 \documentclass[twocolumn]{aastex63}

\newcommand{\fagn}{f_{\rm AGN}}

\newcommand{\rtwenty}{\rho_{20}}

\newcommand{\ana}{A\&A}

\newcommand{\kms}{{\rm~km~s^{-1}}}
\newcommand{\kpc}{{\rm h^{-1}~kpc}}

\newcommand{\SFRf}{\rm SFR_{fib}}

\newcommand{\barrho}{\bar{\rho}}

\newcommand{\rp}{r_{\rm p}}
\newcommand{\rvir}{r_{\rm vir, nei}}

\newcommand{\fbar}{\mathinner{f_\mathrm{bar}}}

\received{**}
\revised{**}
\accepted{\today}
\submitjournal{ApJL}

\shorttitle{The relative role of bars and environments in AGN triggering}
\shortauthors{M. Kim et al.}
\begin{document}

\title{The relative role of bars and galaxy environments in AGN triggering of SDSS spirals}

\correspondingauthor{Yun-Young Choi}
\email{yy.choi@khu.ac.kr}
\author{Minbae Kim}
\affiliation{School of Space Research, Kyung Hee University\\ 
Yongin, Gyeonggi 17104, Republic of Korea}

\author[0000-0002-0786-7307]{Yun-Young Choi}
\affiliation{School of Space Research, Kyung Hee University\\ 
Yongin, Gyeonggi 17104, Republic of Korea}

\begin{abstract}
We quantify the relative role of galaxy environment and bar presence on AGN triggering in face-on spiral 
galaxies using a volume-limited sample with $0.02 < z < 0.055$, $M_r < 19.5$, 
and $\sigma > 70 \kms$ selected from SDSS Data Release 7. 
To separate their possible entangled effects, we divide the sample into bar and non-bar sample,
and each sample is further divided into three environment cases of isolated galaxies,
interacting galaxies with a pair, and cluster galaxies. 
The isolated case is used as a control sample.
For these six cases, we measure AGN fractions at a fixed central star formation rate and 
central velocity dispersion, $\sigma$.
We demonstrate that the internal process of the bar-induced gas inflow is 
more efficient in AGN triggering than the external mechanism of the galaxy interactions in groups and cluster outskirts.
The significant effects of bar instability and galaxy environments are found in galaxies with a relatively less massive bulge.
We conclude that from the perspective of AGN-galaxy co-evolution, a massive black hole is one of the key drivers
of spiral galaxy evolution. 
If it is not met, a bar instability helps the evolution, and in the absence of bars, galaxy interactions/mergers become
important. In other words, in the presence of a massive central engine,
the role of the two gas inflow mechanisms is reduced or almost disappears.
We also find that bars in massive galaxies are very decisive 
in increasing AGN fractions when the host galaxies are inside clusters.

\end{abstract}

\keywords{galaxies: active --- galaxies: nuclei --- galaxies: evolution --- galaxies: spiral --- methods: statistical}


\section{Introduction}
Active galactic nuclei (AGNs) are powered by gas accretion 
onto a central supermassive black hole (SMBH) \citep{Lynden1979, Rees1984}.
According to the standard paradigm, 
AGN activity is expected to be closely related to the mechanism of
the gas inflow into the nuclear region.
Given the formation and evolution of spiral galaxies, possible mechanisms are broadly classified into two categories: 
(1) external processes such as galaxy interactions, major mergers 
\citep{Sanders1988, Springel2005, Kim2020a}, and
minor mergers \citep{Roos1981, Hernquist1995}, (2) 
internal dynamical processes such as turbulence of 
the interstellar medium (ISM) in galactic discs 
\citep{Wada2009}, 
stellar wind \citep{Ciotti2007, Davies2012}, 
and tidal torques due to non-axisymmetric perturbations
such as bars \citep{Friedli1993, Athanassoula2003}.

The large volume-limited galaxy sample obtained from the Sloan Digital Sky Survey (SDSS)
makes it possible to conduct many statistical studies that 
compare changes in AGN fractions with different galaxy properties,
bar presence, and various galaxy environments. 
It has been shown that AGN activity is closely related to 
the gas inflow mechanisms of bar instability 
\citep{Oh2012, Alonso2013, Galloway2015, Kim2020b}, 
galaxy interactions/mergers \citep{Ellison2011, Hwang2012, Goulding2018, Kim2020a},
and high-local density environments \citep{Gilmour2007, Sivakoff2008, Pimbblet2012, Sabater2013}.

These investigations are essential for understanding 
galaxy formation and evolution, but
there is still no clear understanding of the observed galaxy-AGN co-evolution.
Here, we focus on the fact that external processes affect physical quantities
\citep{Blanton2005a,Park2009a,Park2009b,Li2019} and 
bar presence \citep{Eskridge2000, vandenBergh2002, Lee2012a}, and 
ultimately control the central gas supply to the galactic center
\citep[e.g.][]{Sabater2015}. 
The entanglement between these three primary factors may have provoked 
conflicting observational results that depend on sample selection.

In this study, we aim to investigate the direct link
between AGN and galaxy evolution after 
minimizing the possible entanglement by dividing the
sample into sub-samples that can allow as to isolate 
the effect of each of the three primary factors.
To this end, 
we first divide the sample into two categories - barred and non-barred.
Then, each category is sub-divided into three subsets based
on the environment - isolated, groups, and clusters.
Thus, we obtain a total of six sub-samples.
For each sub-sample, we measure AGN fractions at a fixed central velocity dispersion and central star formation rate (SFR) and compare the results.
Finally, we quantify the relative roles of the gas transport mechanisms.

\section{Observational Data and Sample Selection}
\label{sec:sample}
We select a volume-limited sample with an $r$-band absolute magnitude $\mathrm{M_{r}<-19.5}$ 
and redshifts $0.020<z<0.055$ selected from the Sloan Digital Sky Survey Data Release 7 \citep[SDSS DR7;][]{Abazajian2009}. 
The fiber star formation rates, $\SFRf$ is 
obtained from the Max Planck Institute for Astrophysics and the Johns Hopkins University (MPA/JHU) DR8 catalog  \citep{Brinchmann2004}. 
The stellar velocity dispersion, $\sigma$ is adopted 
from the New York University Value-Added Galaxy Catalog \citep{Blanton2005b}.
A simple aperture correction of the $\sigma$ is made using the formula of \citet{Bernardi2003}.
Here, we use only galaxies with $\sigma>70 \kms$ to avoid a selection effect due to the [O III] flux limit in detecting the AGN. 
The $\sigma$ and $\mathrm{M_{r}}$ cuts exclude many disk-dominated and irregular late-type galaxies in our sample.

\subsection{Morphology Classification}
We also limit the galaxy sample to spiral galaxies by adopting the morphological 
classification of the Korea Institute for Advanced Study DR7 Value-Added Galaxy Catalog 
\citep[KIAS DR7-VAGC;][]{Choi2010}. 
They adopted an automated classification scheme introduced by \citet{Park2005} 
and corrected misclassifications due to the automated scheme 
by an additional visual inspection.

Then we classify barred galaxies by adopting the barred galaxy catalog provided by \citet{Lee2012a}.
They defined barred galaxies as galaxies with a bar size larger than approximately 
$25\%$ of the galaxy size by visual inspection.
Since the bar fraction is affected by the inclination of galaxies,
we also limit the late-type galaxy sample to those with an isophotal axis ratio $b/a$ greater than 0.6.

Our final sample consists of 6195 spiral galaxies with $\sigma >70\kms$
and 1893 (30.6\%) are barred and 3754 (60.6\%) are non-barred. The rest are weak or ambiguous barred galaxies.

\subsection{AGN Selection}

Type ${\rm II}$ AGNs are separated from star forming galaxies (SFGs) based on the flux ratios of the Balmer and ionization line \citep[BPT diagram;][]{Baldwin1981}. The activity types are classified based on the ratios of emission lines ($\rm H\alpha$, $\rm H\beta$, $[\rm OIII]\lambda5007$, and $[\rm NII]\lambda6584$) that are detected with a signal-to-noise ratio of ${\rm S/N}\geq3$. 
We classify the activity types of galaxies using a conservative AGN definition from \citet{Kewley2006} and
define an AGN host by combining the composite galaxies and pure AGNs.
Some of the weak LINERas are retired galaxies powered by the hot low-mass evolved stars rather than low-luminosity AGNs \citep{Cidfernandes2010, Cidfernandes2011}. By adopting a criterion of \citet{Cidfernandes2011}, we excluded ambiguous objects with a $W_\mathrm{H\alpha}<3\AA$ from the pure AGNs. We also excluded potential Type {\rm I} AGNs that have a $\rm H\alpha$ emission line width larger than $\sim 500\kms$ (FWHM). 
Out of 1893 barred-spiral galaxies with $\sigma>70\kms$, 805 AGN hosts (42.5\%) are found
and out of 3754 non-barred ones, 1098 AGN hosts (29.3\%) are found.

\subsection{Environmental Parameters}

Two major environmental factors are considered in this study. 
One is a large-scale background density $\rtwenty$ defined 
by the twenty closest galaxies of a target galaxy in the sample. 
This density is measured across distances just over a few Mpc \citep[see Sec. 2.5 of ][for details]{Park2009a}. The other environmental factor is the distance between a target galaxy and a pair galaxy, $\rp$.
Each environment of all the sample galaxies is described by a combination of $\rp$ and $\rtwenty$.
The full details of the estimation of $\rtwenty$ and $\rp$ are described in \citet{Park2008} and \citet{Park2009a}.

\subsubsection{Large-scale Background Density}

The large-scale background density of a target galaxy is given by 
\begin{equation} 
\rho_{20}({\bf x})/{\bar\rho} = \sum_{i=1}^{20} \gamma_i L_i W_i(|{\bf x}_i - 
{\bf x}|)/{\bar\rho}, 
\end{equation} 
where the $\gamma_i$ is the mass-to-light ratio of a background galaxy
that is adopted to obtain the mass density described by 20 neighboring galaxies.
Here, the ratio of the dark halo virial mass for early- and late-type targets,
$\gamma$(early) $= 2\gamma$ (late) is all that is needed. 
The $\bar\rho$ is a mean density of the universe with a total volume of $V$, and 
$L_i$ is the $r$-band luminosity of the closest 20 background galaxies of a target spiral galaxy.
We adopt the spline kernel weighting, $W_i$, that has an adaptive smoothing scale
to include 20 galaxies within the kernel weighting.

\subsubsection{The Nearest Neighbor Galaxy}
\label{sec:sec2.4.2}
The pair galaxy for a host galaxy is defined using 
the conditions of the $r$-band absolute magnitude and radial velocity difference,  $\Delta v$,
as that which is located closest to the target galaxy in the sky.
If a host galaxy has $M_r$, 
the nearest neighbor galaxy for that host galaxy has $M_r < M_r+0.5$ and $\Delta v< 400\kms$, 
making it the most influential neighbor.
The $\Delta v=400\kms$ is
obtained by measuring the pairwise velocity difference between target galaxies and their neighbors 
\citep[see Sec. 2.4 of ][]{Park2008}.
The $\rp$ measures the impact of interactions with the most influential neighbor (i.e., pair galaxy).
The virial radius of the pair galaxy $r_{\rm vir, nei}$ is defined as $\rp$, 
where the mean mass density $\rho_n$ within the sphere with a radius of $r_p$ is equal to 740 times 
the mean density of the universe $\bar{\rho}$.
The $\rvir$ of spiral galaxies with $M_r=-19.5$ corresponds to $210~\kpc$.

\section{Entangled Effects of Environments and Bars}
\label{sec3}
We begin by showing how the probability of a galaxy hosting an AGN or a bar ($\fagn$ or $\fbar$)
varies depending on the two central properties of the velocity dispersion, $\sigma$, 
and central star formation rate, $\SFRf$, 
which are closely related to the BH mass and central SF,
respectively.

\begin{figure*}
    \centering   
   \includegraphics[scale=0.8]{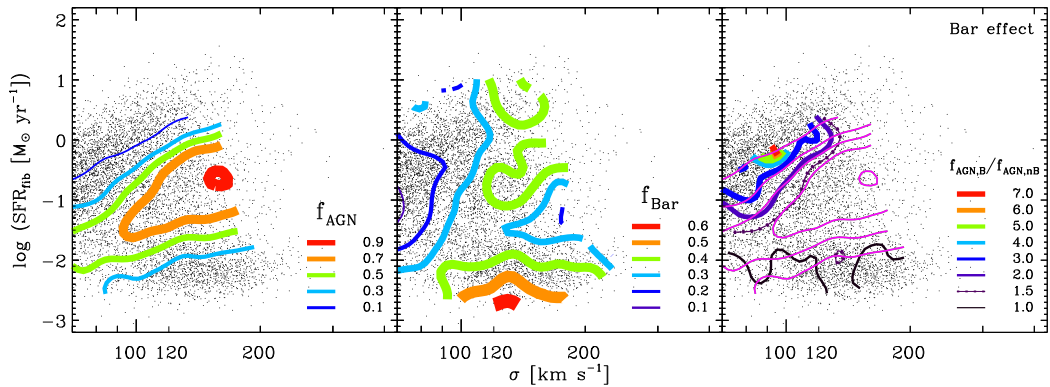}
    \caption{Dependence of AGN fraction (left), bar fraction (middle), and bar effect on $\fagn$ (right)
    on $\SFRf$ and $\sigma$.
    Points are all galaxies of the sample, and contours show constant levels of each measurement. 
    In the right panel, the contours of $\fagn$ are superimposed in magenta for comparison.
}
    \label{fig:fig1}
\end{figure*}

Figure~\ref{fig:fig1} shows how the $\fagn$ and $\fbar$ are related to each other 
at given $\SFRf$ and $\sigma$ in the left-hand and middle panels.
Right panel is for the bar effect on $\fagn$ 
defined as a ratio between the $\fagn$s in the barred 
and the non-barred galaxies. 
A ratio greater than 1.0 indicates a positive bar effect on AGN triggering.
Colored line contours denote the constant levels of $\fagn$, $\fbar$ and bar effect.  

All the smoothed distributions that we measure hereafter are 
obtained using the fixed-size spline kernel for each bin (60 by 60) in the parameter space 
and contours where a standard error estimated by 1000 bootstrapping sampling is more than 30\% 
of the fraction measurement are eliminated.

The key results are as follows.
We have examined the relations
using conservatively selected AGNs with a ${\rm S/N}\geq6$
in \citet{Kim2020b} (see more details).

\begin {itemize}
\item
At a given $\SFRf$ and $\sigma$, it is clear that overall,
the bar presence has a positive effect on $\fagn$.
However, the $\fbar$ and the bar effect on $\fagn$ are not directly related.
That is, the bar presence itself has little to do with nuclear activity.

\item 
In high-$\sigma$ galaxies having the highest $\fbar$ and $\fagn$,
the bar effect is rather the lowest, 
implying that the BH mass is a key driver of galaxy evolution.

\item
The strongest bar effect is found in SFGs leaving the main sequence 
with high $\SFRf$ and low $\sigma$ values, 
although they have low $\fagn$ and $\fbar$ values.
In galaxies where BH is not massive 
but actively stars form, bars play an important role in inducing AGN.

\end{itemize}

\begin{figure*}
    \centering   
   \includegraphics[scale=0.9]{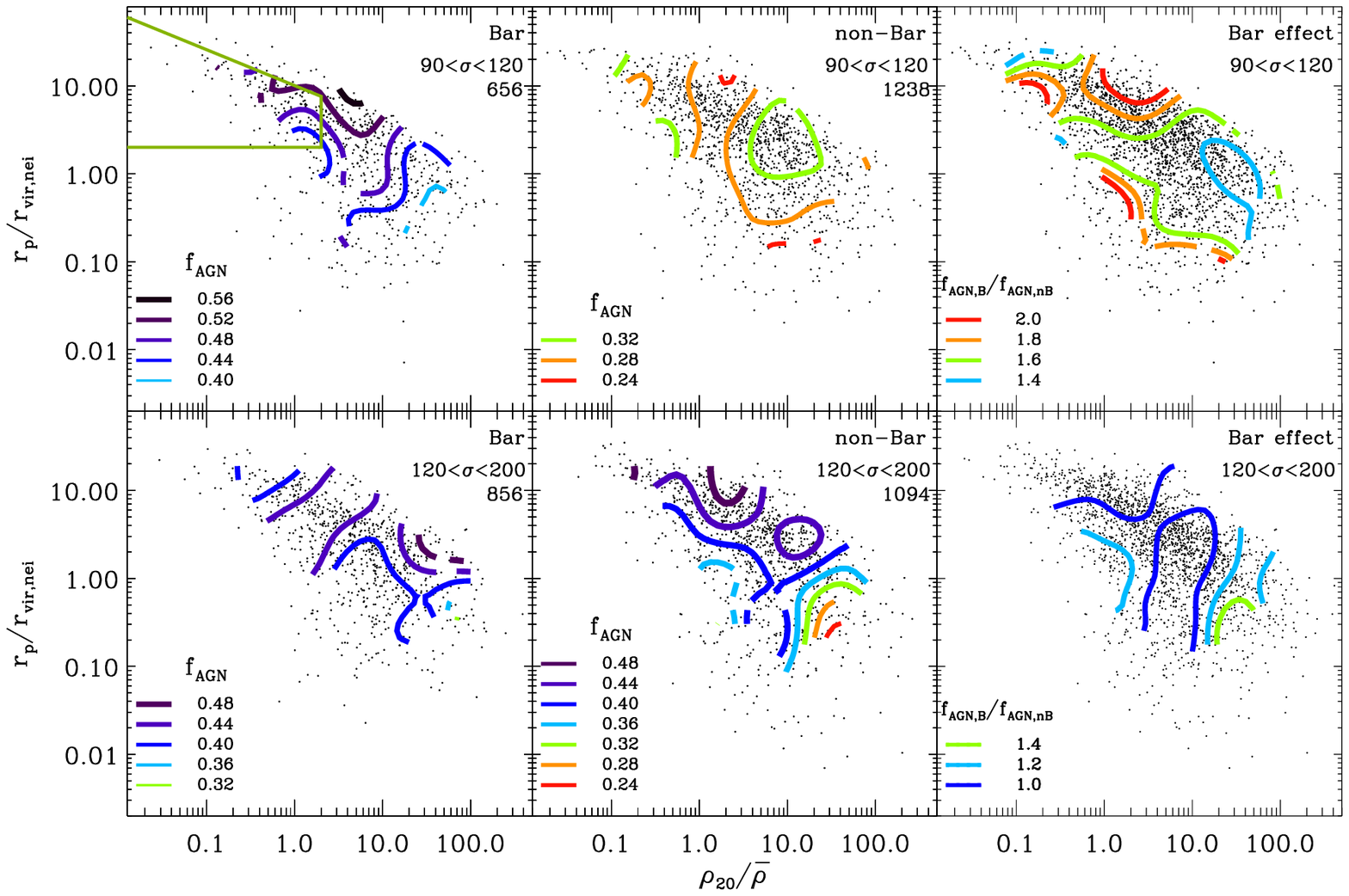}
    \caption{
    AGN fraction and environment relations. Four cases are given: barred galaxies (left)
    and non-barred galaxies (middle) 
    with $90\kms<\sigma<120\kms$ (upper) and $120\kms<\sigma<200\kms$ (bottom). 
    An environment of a galaxy is given with the projected pair separation $\rp$ 
    and the large-scale background density $\rtwenty$. 
    Right panels are for the bar effect on AGN fraction of each sub-sample. 
    The total galaxy number of each sub-sample is given in the corresponding panel.     
    The region enclosed by the green lines in the upper left panel represents the isolated environment.
}
    \label{fig:fig2}
\end{figure*}

Here a natural question arising is whether or not the AGN fraction directly depends on 
environments. To this end, first of all, it is necessary to minimize the effects of the bars and $\sigma$.
We first divide the sample into two cases with $\sigma > 120\kms$ and $90<\sigma<120\kms$.
The lower cut of the $\sigma = 90 \kms$ is to avoid possible incompleteness.
Each sample is further divided into two, according to the presence or absence of a bar.
Then we measure $\fagn$ in the $\rp$-$\rtwenty$ space for the four sub-samples,
seen in Figure~\ref{fig:fig2}.

The $\rp$-$\rtwenty$ diagram
well describes the various environments in which galaxies reside
\citep[see Sec. 3.2.1. of][for details]{Kim2020a}.
The $\rtwenty$ spans 
various large-scale environments from voids to clusters and 
the $\rp$ measures the impact of interactions with the most influential neighbor.

For convenience, we define 
a intermediate-local density region of $\barrho<\rtwenty<10\barrho$ 
and a high-local density region of $\rtwenty >20\barrho$
as a group environment and cluster environment, respectively.

\begin{figure*}
    \centering   
   \includegraphics[scale=0.9]{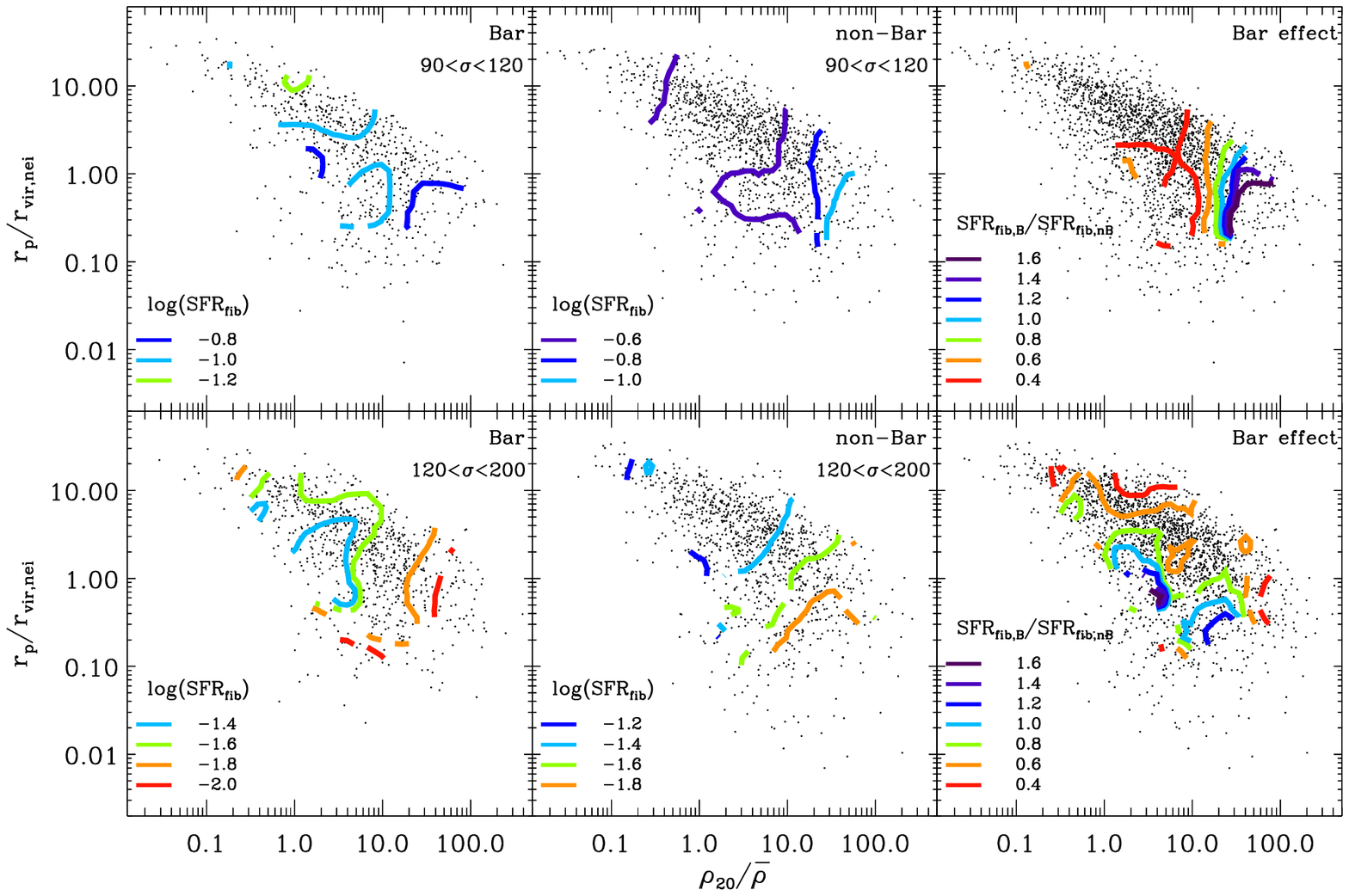}
    \caption{ Median contours of $\SFRf$ in the $\rp$-$\rtwenty$ space 
    for the four cases (same as in Fig.~\ref{fig:fig2}). 
    Contours are limited to regions with statistical significance above 1-$\sigma$. 
    Right panels are for the bar effect on $\SFRf$ of each sub-sample. 
    A smaller ratio than 1.0 indicates a positive bar effect on SF quenching.
    } \label{fig:fig3}
\end{figure*}

We also present the changes in $\SFRf$ in the same $\rp$-$\rtwenty$ space for the same sub-samples,
seen in Figure~\ref{fig:fig3}. 
Note that the median contours of $\SFRf$ are uniformly binned in the logarithm of $\SFRf$.
In the right panels, the bar effects on $\SFRf$ are presented.  
A smaller ratio than 1.0 indicates a positive bar effect on the central SF `quenching.'
The values of $ \SFRf $ of each $\sigma$ sub-sample are 
somewhat limited due to the $ \sigma $ limit.

Spiral galaxies tend to disappear at smaller $\rp$s in the cluster environment.
Given the morphology-density relation, 
the inner region of clusters seems to be occupied by elliptical galaxies.
A region with a relatively larger $\rp = 1\sim 2\rvir$ corresponds to cluster outskirts.
In the group environment, galaxies at $\rp < \rvir$ hydro-dynamically interact with the closest neighbor,
which is enough to change the mean morphology and SF activity of the target galaxies \citep[e.g.][]{Park2009a}. 

Meanwhile, we classify galaxies that do not possess a close pair (i.e., large $\rp$) 
and are surrounded by few neighbors (i.e., low $\rtwenty$) 
as isolated ones that we use as a control (see Figs.~\ref{fig:fig4} and \ref{fig:fig5} below).
The region enclosed by the green lines in the upper left panel of 
Figure~\ref{fig:fig2} represents the isolated environment.

Figure~\ref{fig:fig2} shows that AGN triggering depends on the environment, which is different depending on the presence of a bar and the $\sigma$ value. This finding demonstrates that, when investigating the direct impact of the environment on AGN, it is necessary to limit carefully galactic properties closely related to the central cold gas supply and bulge mass.

In the upper panels, the low-$\sigma$ sub-sample clearly shows that overall the barred case has a significantly higher $\fagn$ compared to the non-barred counterpart in a given environment, demonstrating a critical role of bars. This fact is consistent with the third key result of Figure~\ref{fig:fig1}.

At the largest $\rp$ in groups, the non-barred case has the highest $\SFRf$ and the lowest $\fagn$, while the barred case has a relatively lower $\SFRf$ and the highest $\fagn$. As a result, the bar effect in the largest $\rp$ regions of groups doubles (see the upper right panel). Previous works \citep{Park2008, Park2009a} pointed out that the galaxies at the location would be end-products of mergers and strong interactions.

At the same location in the upper right panel of Figure~\ref{fig:fig3}, the ratio between the $\SFRf$ in the barred and the non-barred galaxies is smallest (less than 1.0), indicating that the SF quenching in the galactic center is also strongly accompanied by the bars.

These findings demonstrate that bar instability promotes both central SF quenching and BH feeding in galaxies, 
seen at the late stage of gas-rich mergers in groups. 
The anti-correlation between central SFR and AGN activity supports negative feedback \citep[e.g.,][]{Silk1998, Fabian2012}.

A caveat here is that as \citet {Robichaud2017} pointed out in numerical simulations, 
AGN-driven outflow in barred hosts can collide with inflowing gas, 
possibly leading to SF enhancement in the central kpc region (i.e., positive plus negative scenario). 
Indeed, in \citet{Kim2020b} using the same low-$\sigma$ sample, 
we found a tendency that at a given $\SFRf$ and $\sigma$, barred cases of AGN hosts 
have a relatively stronger outflow signature (traced by [O III] velocity dispersion) than non-barred counterparts. 
In the high-$\sigma$ sample, the tendency is rare. 
Assuming that strong outflow of AGNs in barred galaxies causes SF enhancement along with SF quenching, 
barred AGN hosts could have the same central SFR as that of the non-barred ones.

However, the most pronounced anti-correlation was found in the barred samples of this study, 
especially in galaxies that have experienced a recent gas-rich merger.
Probably due to violent disturbances caused by gas-rich mergers,
bars drive larger amounts of gas toward the center, increasing both accretion rate and AGN outflow strength. 
This feature suggests that the positive feedback effect is not sufficient to mitigate the negative feedback effect, at least within the central region. This study shows that barred galaxies with less massive BH are excellent samples 
to understand the relation between AGN feedback and galaxy evolution.

Meanwhile, in the lower panels, the high-$\sigma$ sub-sample shows a little bar effect overall. The feature is because galaxies with a massive bulge can have high $\fagn$ without bars. The result reveals that the BH mass is a crucial driver of galaxy evolution, consistent with the second key result of Figure~\ref{fig:fig1}.

A noticeable bar effect is observed in clusters. Compared to the barred ones, non-barred ones have a sharply decreasing $\fagn$ towards the center. In other words, in galaxies with a massive BH, bar-driven gas inflows are useful only in clusters where there is a deficit of available cold gas fuel. They tend to have a less luminous AGN and red color 
\citep[][for details]{Kim2020a}.
\citet{Alonso2014} found that the enhancement of nuclear activity is notable in barred active galaxies located in higher-density environments using a massive galaxy sample, consistent with our result.

\section{The relative Importance of a massive central engine, bars, and galaxy interactions}\label{sec4}
In the previous section, we found in non-barred galaxies that the environmental dependence of $\fagn$ exists even after excluding the effects of bars and bulges. However, their $\SFRf$ value still has an environmental dependency, which can affect $\fagn$ (in particular, the low-$\sigma$ case).

Therefore, to properly remove the dependence on the primary central quantities, we measure the $\fagn$ values at given $\SFRf$ and $\sigma$. We divide the sample into three different environment cases. Each case is further divided into two cases, those with and without a bar.

\begin{figure*}
    \centering   
   \includegraphics[scale=.8]{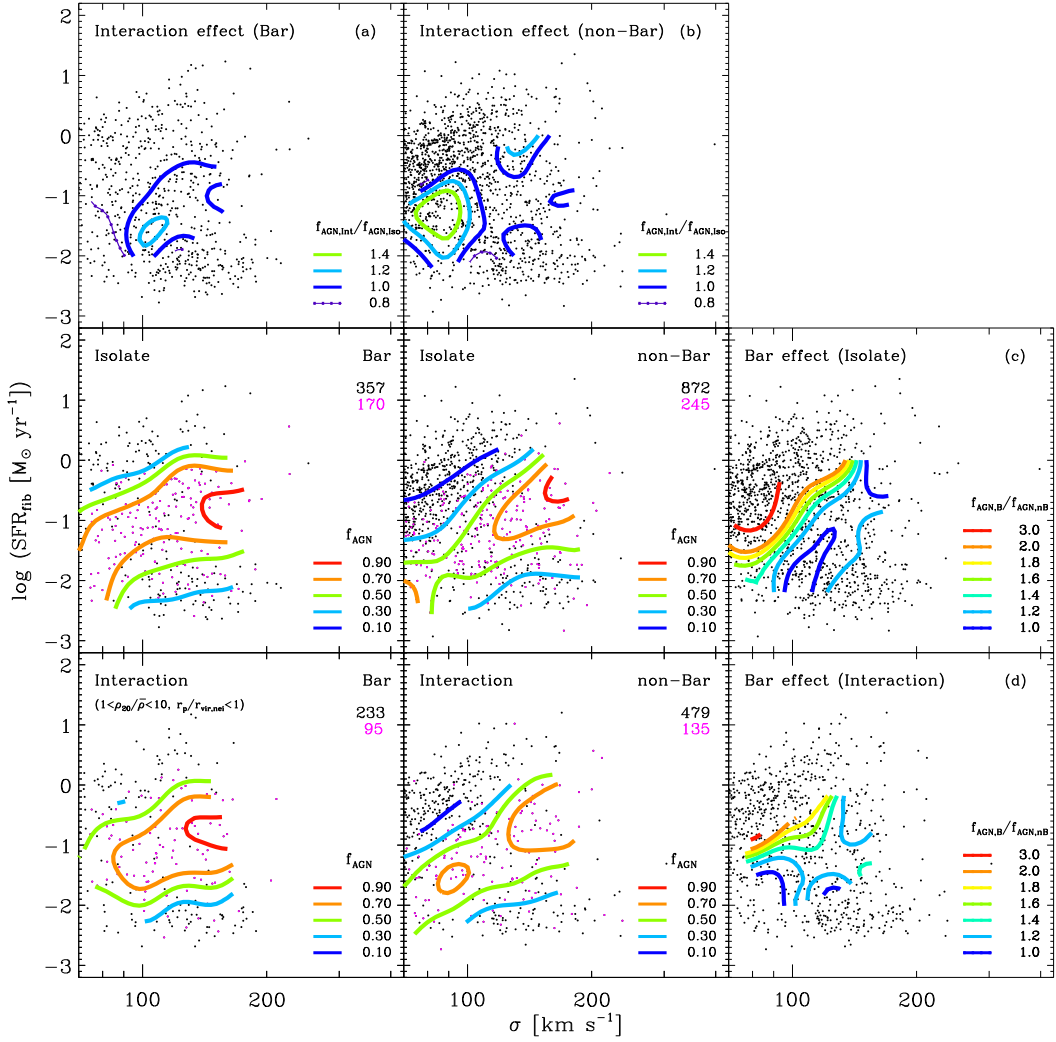}
    \caption{Dependence of $\fagn$ on $\SFRf$. 
    Four cases are given: barred- and non-barred cases located in isolated environments (middle), 
    and barred- and non-barred cases interacting with a pair (bottom). 
    The relative effect of a pair interaction is plotted for the barred (panel $\it a$) 
    and non-barred cases (panel $\it b$). 
    The isolated cases are used as a control. 
    Panels $\it c$ and $\it d$ are for the bar effect in a given environment. 
    In each sub-sample, magenta and black points are AGN hosts and non-AGN galaxies, 
    respectively and their total number is also given in the same color as the points.
    }\label{fig:fig4}
\end{figure*}

\begin{figure*}
    \centering   
   \includegraphics[scale=.8]{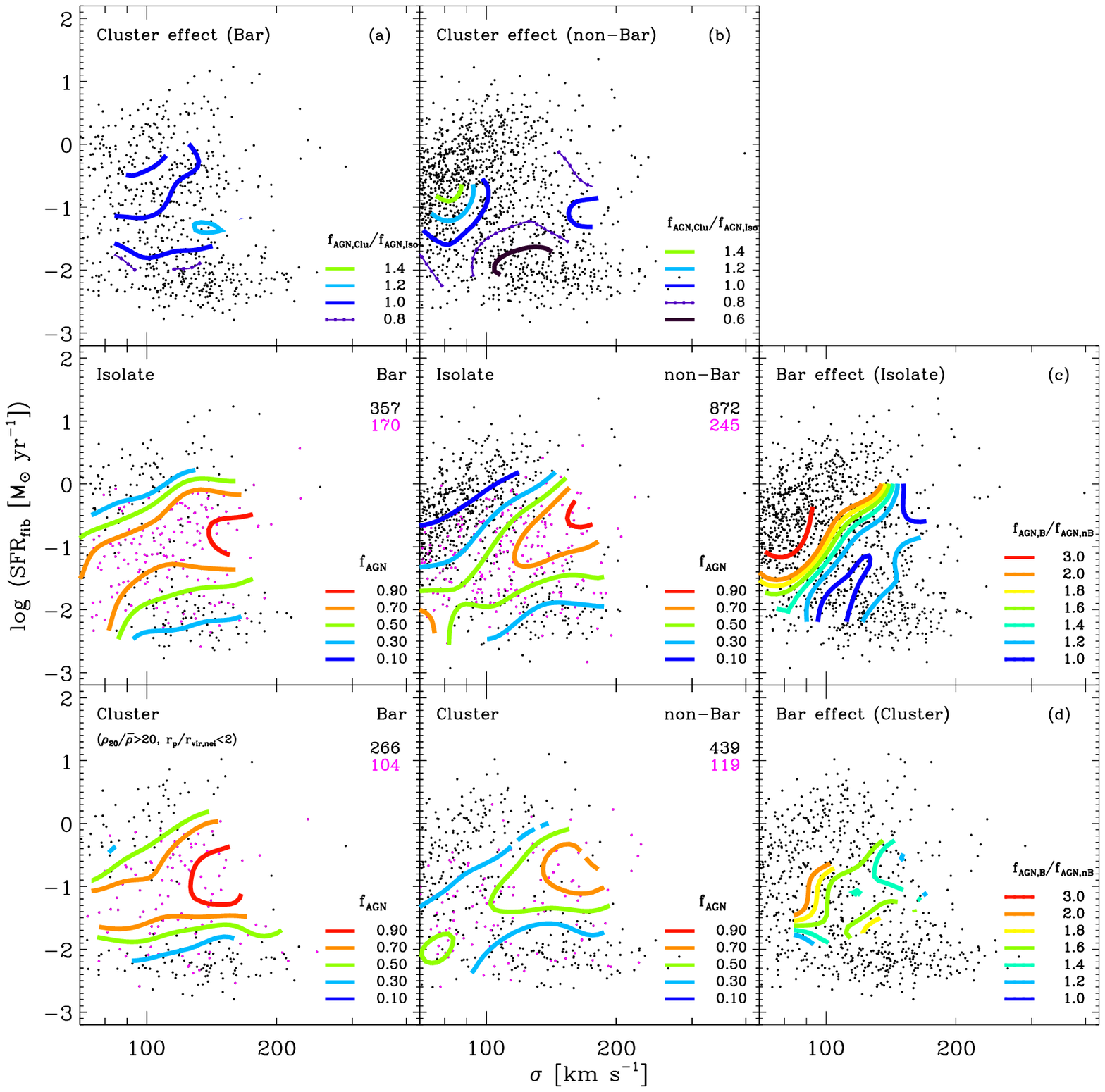}
    \caption{Same as in Fig.~\ref{fig:fig4}, but for the relative effect of cluster environments.}
    \label{fig:fig5}
\end{figure*}

The results for a total of six sub-samples are shown in Figures~\ref{fig:fig4} and \ref{fig:fig5}. Besides, we measure the relative effects of the pair interaction and cluster environment against the isolated environment for barred and non- barred samples, separately. We also measure the bar effect in a given environment. The panel $b$ in each figure (i.e., the non-barred case) shows the impact of the environment only, and the panel $c$ in each figure (i.e., the isolated case) shows the bar effect with a minimal environmental contribution.

The $c$ and $d$ panels in Figures ~\ref{fig:fig4} and \ref{fig:fig5} demonstrate that bars play the most crucial role when SFGs with lower $\sigma < 100\kms$ evolve to the starburst-AGN composite hosts, which most favors isolated environments. Conversely, for high-$\sigma$ galaxies with the largest $\fagn$, the bar effect is the smallest. For gas-poor galaxies hosting faint AGN, the bar effect resumes, which is more noticeable in dynamic environments, especially a cluster environment.

In the $b$ panels of Figures~\ref{fig:fig4} and \ref{fig:fig5}, non-barred galaxies show the $\sigma$ dependency of the environmental effect. The decisive role of galaxy interactions or cluster environments compared to an isolated environment is only observed at low $\sigma$s. At high $\sigma$s, the low value of the bar effect shows how harsh a cluster environment is for AGN triggering.

By comparing the $a$ and $b$ panels in Figures~\ref{fig:fig4} and \ref{fig:fig5}, we infer that once a galaxy has a bar, $\fagn$ enhances overall and is less affected by an environment. The $\sigma$ dependency of the environmental effect is also not as evident as in non-barred galaxies overall.

By comparing the $b$ and $c$ panels in Figures~\ref{fig:fig4} and \ref{fig:fig5}, we quantify the relative role of a bar and environments. The bar-induced gas inflow is approximately 20\% to 2 times more efficient at AGN triggering than the external mechanisms. Even the importance of the bar effect tends to be more significant at lower $\sigma$s. 
\citet{Alonso2018} suggested that barred AGN hosts show an excess in AGN activity and BH accretion rate compared to AGN hosts with a close pair using a massive galaxy sample, consistent with our result.

We conclude that environmental factors of galaxy interaction or cluster environments play a decisive role in low-$\sigma$ galaxies. Once there is a large bar in a galaxy, the environmental factors have little impact on AGN triggering. We find the most substantial bar effect when the low-$\sigma$ galaxies are in isolated environments.

Galaxies with a massive bulge have a high $\fagn$ even without a bar. However, for the red galaxies that have consumed cold gas fuel, the role of the bar becomes critical again, especially when in clusters.

\section{Summary}
Using an extensive volume-limited sample of spiral galaxies obtained from the SDSS DR7, we showed that the AGN fraction of galaxies is quantitatively different depending on the central velocity dispersion and central SF 
of the host galaxy, bar presence, and galaxy environments. 

We found that when galaxies with a massive BH have high central SFRs, 
AGNs are best triggered, showing that the BH mass is the most crucial driver of spiral galaxy evolution. 
For galaxies with a less massive BH or galaxies with low central SFRs due to lack of central gas supply, 
even with a large BH mass, bar instability plays a vital role in galaxy evolution. 
We found the most substantial effect of bars on AGN in SFGs (i.e., blue galaxies) 
evolving to AGN host galaxies, consistent with the results of previous studies \citep{Hao2009,Oh2012}.

We also investigated whether galaxy environments providing other gas inflow mechanisms directly affect AGN activity in the innermost part. Since galaxy environments directly affect the critical ingredients for AGN triggering, such as BH mass, gas fuel, and bar formation, to unveil the direct impact of the galaxy environment, the properties should be carefully limited. Indeed, the role of galaxy environments has often been debated. 

The combination of pair interactions and local density well describes the various environments of galaxies, allowing relative comparisons between different environments at a glance. We successfully isolated each effect of bars and galaxy environments. 

In particular, this study highlights how directly galaxy environments influence AGN-galaxy co-evolution. Gas-inflows induced by bars or galaxy environments play a decisive role in BH feeding of galaxies with a less massive BH.  In the absence of the massive central engine or gas fuel availability, the role of the additional gas inflow mechanisms becomes critical. 

\acknowledgments
The authors thank Sungsoo S. Kim for helpful comments.
This research is supported by the National Research Foundation (NRF) of 
Korea to the Center for Galaxy Evolution Research (No. 2017R1A5A1070354). 
MBK is also supported by the BK21 plus program through the NRF funded by the Ministry of Education of Korea.

Funding for the SDSS and SDSS-II has been provided by
the Alfred P. Sloan Foundation, the Participating Institutions, the National Science Foundation, the U.S. Department of Energy, the National Aeronautics and Space Administration, the Japanese Monbukagakusho, the Max Planck Society, and the Higher Education Funding Council for England. The SDSS Web site is http://www.sdss.org/.

\bibliographystyle{aasjournal}



\end{document}